# SciDataFlow: A Tool for Improving the Flow of Data through Science

September 7, 2023

**Vince Buffalo**

vsb@berkeley.edu

Department of Integrative Biology,

University of California, Berkeley

## Abstract

Managing data and code in open scientific research is complicated by two key problems: large datasets often cannot be stored alongside code in repository platforms like GitHub, and iterative analysis can lead to unnoticed changes to data, increasing the risk that analyses are based on older versions of data. Here, I introduce *SciDataFlow*: a fast, concurrent command-line tool paired with a simple Data Manifest specification. SciDataFlow streamlines tracking data changes, uploading data to remote repositories, and pulling in all data necessary to reproduce a computational analysis.

## Introduction

The complexity of modern computational scientific projects can be immense. In genomics, project directories can contain terabytes of raw and intermediate data scattered across numerous files. These files often reside alongside processing scripts and analysis code that refines data into the outputs of a computational analysis: figures, summary statistics, supplementary files, and ultimately scientific knowledge. While simply organizing and managing large computational projects is already an arduous task, two problems further complicate computational research.

First, while code is now often managed using version control systems like Git and hosted on platforms like GitHub (Ram 2013, Buffalo 2015), the data —often too large to be hosted on these sites— is often stored on data repository sites like FigShare, Zenodo, or Dryad. Unfortunately, this *data-code divide* bifurcates scientific data from the code that produced it (i.e. output data) or operates on it (i.e. input data). Reuniting the data associated with a project would entail downloading and placing it within the locations expected by the computational pipeline. Overall, this is time-consuming, limits computational reproducibility, and is a fragile approach that relies on scientists to carefully document where data is within a project.

Second, computational projects often involve rerunning data processing steps or analyses several times, as software bugs are corrected, input data is changed by upstream steps or collaborators, or computational methods are improved. While new software has helped resolve complicated dependency issues in computational workflows (Köster and Rahmann 2012), the dynamic evolution of a computational project often leads the researcher to ask, "Which data has changed?". Since scientific results are conditional on data produced at various steps, this *silent data modification* issue can lead to erroneous results if the researcher is not alerted when data is changed.

Here, I introduce *SciDataFlow*, which is both a (1) simple specification for a *Data Manifest* used to track the data in a project, and (2) a fast, concurrent command-line tool to register data in the Data Manifest, track when data changes, retrieve data from and upload data to external repositories, and store important metadata in a project. The SciDataFlow tool `sdf` supports both FigShare and Zenodo, and is designed to be extensible so that other data repository services with REST interfaces (Fielding 2000) can be added in the future.

### The Data Manifest Specification

The Data Manifest format is a simple, human and machine-readable specification in the YAML format (Listing 1). It is designed to be minimal, with three primary components. First, there is a `files` section containing file sizes and MD5 checksums of the data versions presently "registered" in the manifest. This information is needed to track when local or remotely stored data has been modified from this registered version. Second, there is a `remotes` section containing the minimal sufficient information to upload and download data from remote data repository services. While data repository services require secret authentication tokens to access their interfaces, this private information is automatically stored and managed outside of the manifest in a user's home directory.

```yaml
files:
  data/supplement/figure_1.tsv:
    tracked: true
    md5: 87c1148fa71abf01daceb82d8fbfee53
    size: 993
remotes:
  data/raw: !ZenodoAPI
    name: ancient_dna_analysis
    deposition_id: 8271457
    bucket_url: https://zenodo.org/api/files/558014a8-8e04-4a7e-b1c9-7c82bcbe8fa9
metadata:
  title: An analysis of new Ancient DNA Samples
  description: This project contains the code and data to reproduce Joan et al. (2023).
```

Listing 1: A small example YAML Data Manifest.

Third, the Data Manifest contains a `metadata` section for project-wide metadata such as a description. Some additional metadata about the local user (e.g. their name, email, affiliation) can also be stored in the user's home directory. All such metadata is automatically propagated to data repositories that support these metadata fields.

Since SciDataFlow does not manage multiple data versions or backups (which would increase complexity and disk space usage), the manifest is plaintext and designed to be tracked by Git. Thus, a simple emergent feature arises: each Git commit also stores the MD5s of data registered in the manifest at that time. This allows the user to leverage Git's powerful history system to track older changes to data.

### The SciDataFlow Command Line Tool

The SciDataFlow command-line tool `sdf` is designed to have an easy-to-use Git-like interface. A new Data Manifest can be initialized in a project directory with `sdf init`. A common workflow would be to register a new data file in the manifest (`sdf add`), indicate this file should be *tracked* by the remote (`sdf track`), link a remote service to the manifest (`sdf link`), and push all tracked files to the remote:

```
$ sdf add data/supplement/figure_1.tsv
$ sdf track data/supplement/figure_1.tsv
$ sdf link zenodo <access_token> --name ancient_dna_analysis
$ sdf push
```

Another common workflow is to clone an existing research project Git repository and use SciDataFlow to reunite this project with its data. This can be accomplished with one command: `sdf pull`. This will concurrently download all files in the manifest into their proper locations within project subdirectories.

Finally, some data in a project may originate from static URLs. For example, most reference genomes or annotation data can be downloaded from static links from Ensembl or the UCSC Genome Browser (Cunningham et al. 2022, Nassar et al. 2023). SciDataFlow supports a single subcommand to download and register files from URLs: `sdf get <link>`. Since large bioinformatics projects may have hundreds of files that need to be programmatically downloaded, SciDataFlow also has the `bulk` subcommand, e.g. `sdf bulk downloads.tsv --column 2 --header`, which would download and register all files in the second URL column of this tab-separated value file, ignoring the first header column.

### Discussion

A core goal of SciDataFlow is to establish new reusable computational workflow patterns and discourage redundant, isolated efforts. For example, small tasks like converting a recombination map to a new genome version or computing summary statistics on a gene set are often unnecessary and independently duplicated among research groups. In addition to wasting valuable time, independent reimplementations introduce multiple points of error and prevent the cumulative improvement of these important scientific data assets. By making it effortless to reunite data with code, SciDataFlow promotes modular and Git-tracked codebases for these tasks, fostering data reuse and improvement by research communities.

## Availability

SciDataFlow is available at https://github.com/vsbuffalo/scidataflow, or installable through the Rust Programming Language's crate system with `cargo install scidataflow`.